\documentclass[12pt]{iopart}

\usepackage{graphicx}
\usepackage{citesort}
\usepackage{rotating}
\usepackage{amssymb}

\begin{document}

      \title[PIC Simulations of the Weibel Instability: Perpendicular
        mode] {PIC Simulations of the Temperature Anisotropy-Driven
        Weibel Instability: Analyzing the perpendicular mode}
      \author{A Stockem\footnote[7]{Now at  GoLP/Instituto de Plasmas e Fus\~{a}o Nuclear, Instituto Superior T\'ecnico, Lisbon, Portugal}, M E Dieckmann and R Schlickeiser}
      \address{Institute of Theoretical Physics IV, Faculty of Physics
        and Astronomy, Ruhr-University Bochum, D-44780 Bochum,
        Germany}
      \ead{anne@tp4.rub.de}

\begin{abstract}
         An instability driven by
         the thermal anisotropy of a single electron species is investigated in a 2D particle-in-cell (PIC) simulation. This instability is the one considered by Weibel and    
         it differs from the beam driven filamentation instability. 
         A comparison of the simulation results with analytic theory provides similar exponential growth rates of the 
         magnetic field during the linear growth phase of the instability. We observe in accordance with previous works 
         the growth of electric fields during the saturation phase of the instability. Some components of this electric          
         field are not accounted for by the linearized theory. A single-fluid-based theory is used to determine the 
         source of this nonlinear electric field. It is demonstrated that the magnetic stress tensor, which vanishes in a 1D geometry, is more important in 
         this 2-dimensional model used here. The electric field grows to an amplitude, which yields a force on the electrons that is comparable
         to the magnetic one. The peak energy density of each magnetic field component in the simulation plane agrees 
         with previous estimates. Eddy currents develop, which let the amplitude of the third magnetic field component
         grow, which is not observed in a 1D simulation.

\end{abstract}


\maketitle

\section{Introduction}\label{intro}
         Plasma instabilities that result in the growth of magnetic fields from noise or in their amplification
         are important for the magnetization of the upstream region of supernova remnant shocks \cite{Bell} or 
         for the cosmological magnetic field generation \cite{sch2}. Two kinetic instabilities, which are based 
         on idealized electron distributions, are frequently discussed in the context of providing the seed fields 
         for further instabilities. These are the Weibel instability and the filamentation (beam-Weibel) instability.
         
         In a plasma with a thermally anisotropic electron distribution, in which the electron thermal spread in one 
         direction is larger than that in the other two, electromagnetic fluctuations (noise) are amplified by the 
         thermal anisotropy-driven Weibel instability (TAWI). A magnetic field is generated, which is coherent 
         on electron skin depth scales. This plasma system is stable against electrostatic instabilities, as long as
         there is no drift between the electrons and ions. The filamentation (FI) or beam-Weibel instability
         \cite{yoo,tau,ach,pet,sil1,sto1,med} acts in systems with counter-streaming electron beams and it can be
         considered as an extreme form of the TAWI. The FI is efficient in plasmas with relativistic streaming
         velocities between electron beams with a comparable density, because the growth time of the 
         electromagnetic instability is small compared to that of the electrostatic instabilities. Recently, the 
         importance of the TAWI and the FI has also been recognized in the fast ignition processes for the inertial
         confinement fusion \cite{kar}.

         Instabilities driven by a thermal anisotropy have been widely examined numerically and analytically
         \cite{mor,lem1,lem2,boro,kaa09,pal2009,stodie} and they are considered also in our present work. The 
         thermal velocity of the bi-Maxwellian distribution 
         is larger along one axis than in its perpendicular plane. This anisotropy induces higher micro-currents 
         along the hotter direction and their magnetic repulsion separates in space the electrons with oppositely 
         directed velocity vectors. The low thermal energy in the perpendicular plane cannot work against the 
         structure formation in this plane. A net current and electromagnetic fields develop. The magnetic energy 
         density can reach in extreme cases up to $1/12$ of the total energy density \cite{lem1,lem2,boro} and it 
         exceeds by far the electric one. The focus of previous studies has thus been on the magnetic field. The 
         neglect of the electric field is probably justified, if the thermodynamic properties of the plasma are
         considered. However, if the electron speeds are well below the speed of light, the electric forces on 
         individual electrons may not be small compared to the magnetic ones. This has motivated several recent
         investigations of the nonlinearly driven electric field. Vlasov and PIC simulations \cite{stodie,kaa09,pal2009}
         have demonstrated, that the electric and magnetic field structures are linked. It turns out that the driver of
         the electric field is the pressure gradient force of the self-generated magnetic field, if the wave spectrum 
         is limited to one dimension \cite{stodie}. 
         
         The magnetic pressure gradient force and the magnetic tension force are both a consequence of the $\mathbf{J}
         \times \mathbf{B}$ force of the self-generated magnetic $\mathbf{B}$-field on the driving current $\mathbf{J}$.
         Only the magnetic pressure gradient force can, however, develop in the system with a one-dimensional wave 
         spectrum investigated in Ref. \cite{stodie}. The magnetic pressure gradient force remains stronger than that 
         due to the magnetic tension as we go from 1D to 2D simulations of the FI \cite{die09}, but it is unknown if 
         this is also true for the TAWI. We address this issue here. We consider immobile ions and a bi-Maxwellian 
         electron distribution with a large temperature along one axis (the parallel component in the following) and 
         a lower temperature in the perpendicular plane. The plasma parameters are similar to those in Ref. \cite{stodie},
         but here the simulation geometry gives rise to a two-dimensional wave spectrum and, thus, to a magnetic tension
         force. We can determine the relative importance of both components of the $\mathbf{J} \times \mathbf{B}$-force 
         for the electric field generation by their direct comparison. We may also expect consequences of the altered
         filament dynamics in a 2D simulation. The magnetic field in a 1D geometry eventually becomes strong enough to 
         keep filaments separated, suppressing their further merging. A 2D geometry allows repelling filaments 
         (oppositely directed current) to move around each other and continue to merge with attractive filaments.
         
         Often, the magnetic trapping 
        mechanism is invoked to explain the saturation of the instability, i.\,e.\ the condition for saturation 
        is given, when the magnetic bounce frequency \(\omega_B = |qkvB_k/mc|^{1/2}\) is of the same order as 
        the growth rate of the TAWI \cite{dav02}. This magnetic trapping mechanism does, however, not take into 
        account the electric field.        
        It is, however, becoming increasingly evident that the electric forces can not be neglected, when the 
        TAWI or the FI saturate \cite{cal02}. The pressure gradient of the magnetic field driven by the FI of 
        counter-propagating electron beams accounts for the electric field in 1D and 2D simulations \cite{row07,die09}.
        This electric field is driven by the magnetic pressure gradient force, if the wave spectrum of the TAWI is 
        one-dimensional \cite{stodie}. Early 2D PIC simulations of the TAWI \cite{mor} did not have the signal-to-noise
        ratio that is necessary to determine the exact source of the electric field \cite{mor} and the recent 
        simulations by \cite{stodie,pal2009,kaa09} did not resolve the 2D wave spectrum of the TAWI. It is thus not 
        well-understood, which mechanism produces the electric fields if the wave spectrum of the TAWI is 
        two-dimensional. For this purpose the spatio-temporal evolution equation of a single-species fluid is considered and the coupling between the perpendicular and parallel components of the magnetic and electric field is analyzed.
         
         The linear theory and the numerical method are outlined in section \ref{ana} and the simulation parameters 
         are specified. In section \ref{numres} the numerical results are presented, where we get
         a similar power spectrum to that observed for the FI \cite{dlsd}. The magnetic energy density exceeds by far the electric energy 
density in our simulation. However, the magnitudes of the electric and magnetic 
forces, which act on the non-relativistic electrons, are comparable; both are 
thus equally important for the particle dynamics. Most importantly, we find that the magnetic
         pressure gradient force by itself is too weak in the 2D simulation to explain the observed electric field. 
         The electric force on the nonrelativistic plasma electrons is instead comparable to that expected from the superposition of the magnetic 
         tension force and the magnetic pressure gradient force. The spatial distributions of the fields are alike but not
         identical. In section \ref{discussion} the results and their implications are discussed.

\section{The instability, initial conditions and the simulation method}\label{ana}
\subsection{The linear instability}
        The Weibel instability is investigated in a homogeneous, collisionless plasma with the initial 
        magnetic and electric field strengths \(\mathbf B_0 = \mathbf E_0 = 0\). The ions form an 
        immobile background that compensates the electron charge. 
        The spatially uniform initial distribution of the electrons is given by
        \begin{equation}
	f_0(v_\perp, v_\parallel) = \frac{1}{(2\pi)^{3/2} v_{th\perp}^2
          v_{th\parallel}}
        \exp\left[-\left(\frac{v_\perp^2}{2v_{th\perp}^2} +
        \frac{v_\parallel^2}{2v_{th\parallel}^2} \right)\right],
        \end{equation}
        where \(v_{th\parallel}= \sqrt{k T_\parallel / m}\) and \(v_{th\perp}= \sqrt{k T_\perp / m}\) 
        denote the thermal velocities of the parallel and the perpendicular components, respectively. 
        The Boltzmann constant is \(k\), \(m\) is the electron mass and \(T_\perp\) and \(T_\parallel\)
        are the respective temperatures. 
        
        An instability is driven by a temperature anisotropy \(A=(v_{th\parallel} / v_{th\perp})^2-1 \neq  0\). 
        We choose the large $A = 399$ here and in the simulation (\( v_{th\parallel} = 20 v_{th\perp} \)), so 
        that we get a large growth rate of the instability and a good signal-to-noise ratio for the electromagnetic
        fields in the simulation. This value of $A$ furthermore allows us to test for a large $A$ the finding
        \cite{lem1,boro}, that the maximum average magnetic energy density is almost independent of the initial 
        anisotropy for \(A \gtrsim 25\). Electromagnetic fluctuations with a wavevector in the perpendicular 
        plane are amplified in this case, according to the well-known dispersion relation for the linear phase 
        of the instability \cite{mar1}
        \begin{equation}
				\frac{k^2 c^2}{\omega_{p}^2} + \frac{\sigma^2}{\omega_{p}^2}
        = - \left[1+\frac{1}{2} (A+1) Z'\left(\frac{\imath \sigma}{k
            v_{th\perp}}\right)\right].
        \end{equation}
        \(k\) and \(\sigma\) are the wave number and the associated
        linear growth rate of the growing electromagnetic oscillations
        with a purely imaginary frequency \(\omega=\imath
        \sigma\). The electron plasma frequency is given by
        \(\omega_{p}=(e^2 n/ \epsilon_0 m)^{1/2}\) with electron
        charge \(e\) and electron number density \(n\) and
        \(Z'(\zeta)=-2[1+\zeta Z(\zeta)]\) is the first derivative of
        the plasma dispersion function \(Z(\zeta)= \pi^{-1/2}
        \int_{-\infty}^\infty d t \, \exp(-t^2)/(t-\zeta) \).
        The normalised wave number \(k_{max}c/\omega_{p}=A^{1/2}\)
        determines the upper limit of unstable wave numbers \(0< k <
        k_{max} \). 
        \begin{figure}[ht!]
          \centering \setlength{\unitlength}{0.001\textwidth}
          \begin{picture}(1100,450)(-260,0)
            \includegraphics[width=7cm]{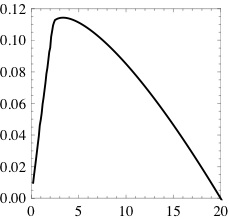}
            \put(-228,-20){\(k_\perp c/\omega_p\)}
            \put(-470,200){
              \begin{rotate}{90}
                \(\sigma/\omega_p\)
            \end{rotate}
            }
          \end{picture}
          \caption{The growth rate \(\sigma(k c/\omega_p)/\omega_p\) as a function of the wavenumber $k$ 
          in the perpendicular plane. After a steep rise at low wavenumbers the maximum is reached. Thermal 
          effects limit the growth rate at large \(k \). The cut-off is given by \(A^{1/2} \approx 20 \).}
          \label{Plot1}
        \end{figure}
\subsection{Numerical method and code resolution}
        The PIC method models self-consistently the interplay of the electric and magnetic fields with a collision-less
        kinetic plasma. The plasma is treated as an incompressible phase space fluid, which is approximated by an 
        ensemble of computational particles (CPs). Each CP has the same charge to mass ratio $q_{cp}/m_{cp}$ as 
        the physical particles it represents. With the relativistic momentum \(\mathbf{p}_{cp}=m_{cp}\gamma
        \mathbf{v}_{cp}\) and velocity \(\mathbf{v}_{cp}\) of a CP, the Maxwell equations for the electric and magnetic
        fields \(\mathbf{E}\) and \(\mathbf{B}\)
        \begin{equation}
	\nabla\times \mathbf E=- \frac{\partial \mathbf B}{\partial
          t}, \, \, \nabla \times \mathbf B= \frac{1}{c^2} \,
        \frac{\partial \mathbf E}{\partial t}+\mu_0 \, \mathbf
        J, \label{thirteen}
        \end{equation}
        and the Lorentz equation for the CP that is located at the position $\mathbf{x}_{cp}$ 
        \begin{equation}
				\frac{\textnormal{d}\mathbf p_{cp}}{\textnormal{d}
          t}=q_{cp} \left ( \mathbf E (\mathbf{x}_{cp})+\mathbf v_{cp} \times \mathbf
        B(\mathbf{x}_{cp}) \right ) \label{fourteen}
        \end{equation}
        are solved. The code fulfills \(\nabla \cdot \mathbf
        E=\rho/\epsilon_0\) and \(\nabla \cdot \mathbf B=0\) to
        round-off precision \cite{eas2}.

        In contrast to the freely moving CPs, the electric and magnetic fields are defined on a grid and have 
        to be interpolated to the position of each CP. With Eq.\ (\ref{fourteen}) the velocity \(\mathbf{v}_{cp}\) 
        is updated and the particle position is advanced in time with \(\textnormal d \mathbf x_{cp} / 
        \textnormal dt = \mathbf v_{cp}\) and the time step \(\Delta_t\). The total current \(\mathbf{J}\) 
        contains the contribution of all microcurrents \(q_{cp} \mathbf{v}_{cp}\), which are then interpolated 
        back onto the grid. Then the electric and magnetic fields are updated with (\ref{thirteen}) and the 
        individual steps are repeated.
\subsection{Initial conditions and simulation setup}
        The parallel axis is aligned with the (unresolved) \(z\) direction (\(v_{\parallel} = v_{z}\)) of the 
        simulation. According to the linear theory, the wave vectors in the \(x\)-\(y\) plane, which we label the
        perpendicular plane, are unstable and $B_x$ and $B_y$ grow exponentially. The rotational symmetry around 
        the parallel direction allows us to combine them to \(B_\perp = B_x + i B_y\). This is allowed if the 
        plasma structures are large compared to a grid cell and small compared to the box, which are both rectangular. 
        The electric field driven through the displacement current is \(E_\parallel = E_z\). In the following, we 
        will discuss the magnetic field in terms of \(c B_\perp\) and \(cB_\parallel \) as they have the same unit 
        as the electric fields \(E_\perp\) and \(E_\parallel \).
        
        The boundary conditions are periodic in all directions. We set the electron plasma frequency to 
        \(\omega_{p}=6.3 \cdot 10^5 \textnormal{ s}^{-1}\). The simulation box is composed of \(N_g \times 
        N_g=1600^2\) rectangular grid cells, each with a side length \(\Delta_x = 10\) m or \(\Delta_x \omega_p 
        / c = 0.021 \) in terms of the electron skin depth, in the \(x\) and \(y\) directions. The total box size
        is therefore \(\left(N_g \Delta_x \omega_p / c \right)^2 = 33.5^2\). The number of CPs per cell is initially
        \(N= 160\). The simulation time step is \(\omega_p \Delta_t\approx 0.0094\) and the simulation time is
        $\omega_p T_{sim} = 434$. The thermal velocities are \(v_{th\parallel}/c \approx 9 \times 10^{-3}\) and
        \(v_{th\parallel} = 20 v_{th\perp}\). 
\section{Numerical results}\label{numres}
\subsection{The energy densities}
\begin{figure}[ht!]
	  \centering
           \includegraphics[width=7cm]{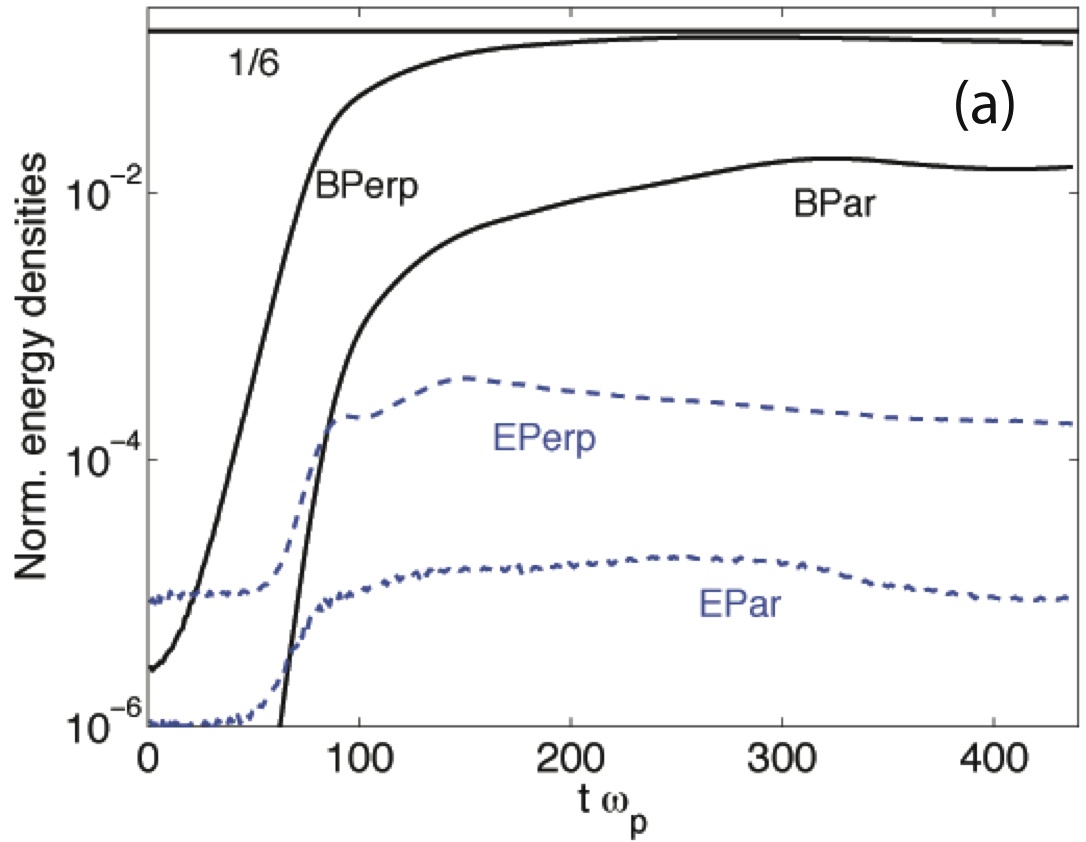}
          \includegraphics[width=7cm]{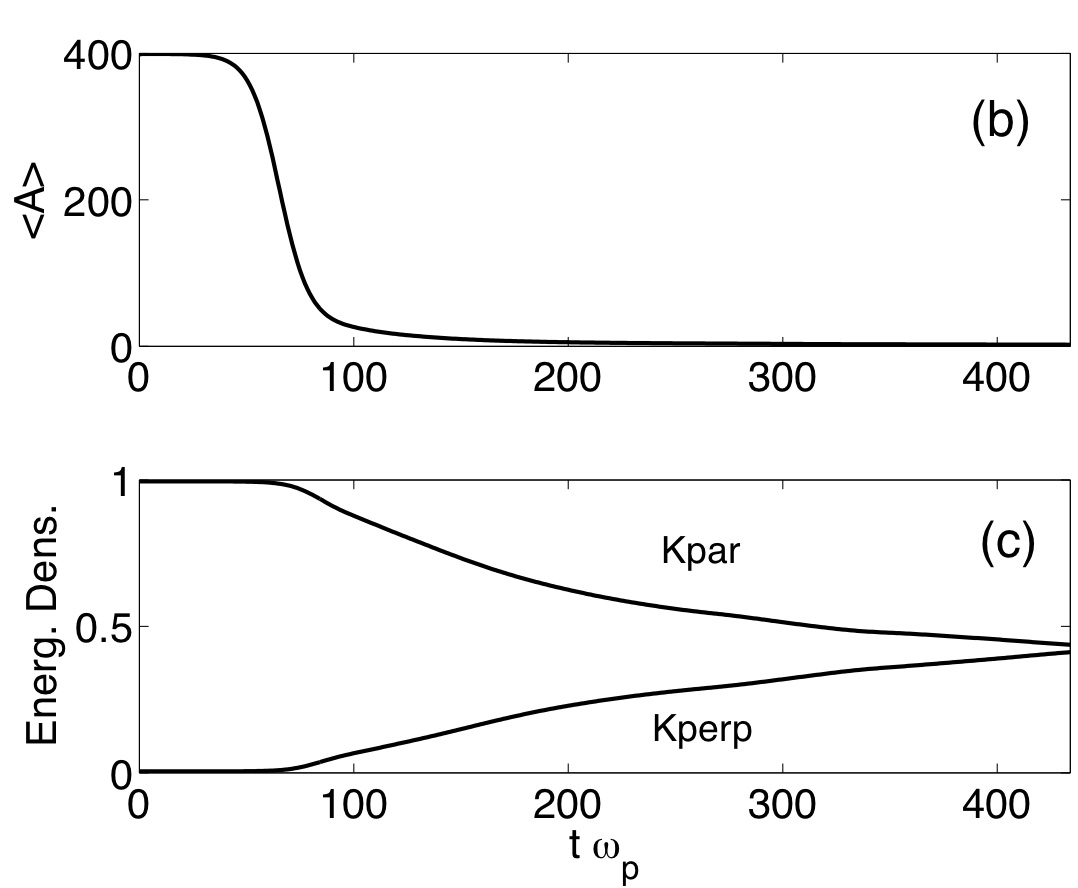}
          \caption{(Colour online) (a) The normalised magnetic energy densities \(\epsilon_{B\perp} /\epsilon_{K0}\) 
          and \(\epsilon_{B\parallel}/\epsilon_{K0}\) (upper black curves) and the electric energy densities 
          \(\epsilon_{E\perp}/\epsilon_{K0}\) and \(\epsilon_{E\parallel}/\epsilon_{K0}\) (lower blue dashed 
          curves). Initially only $\epsilon_{B\perp}$ grows exponentially and the onset of its saturation at 
          $t\omega_p \approx 60$ triggers the growth of the other components. The energy density of $B_\perp$
          peaks above a value 1/12 and just below 1/6 (horizonal line). (b) The decrease of the averaged anisotropy \(<A>\). (c) The kinetic energy density of particles moving in parallel and perpendicular direction.}\label{fig:EnDen}
        \end{figure}

        An investigation of the energy densities provides the overall
        temporal evolution of the instability. Figure \ref{fig:EnDen}
        shows the parallel and perpendicular components of the electric and
        magnetic energy densities, which are given by
        $
        \epsilon_{Es} (t)=N_g^{-2} \sum \limits_{j,k} \epsilon_0
                {|E_s(j\Delta_x,k\Delta_x,t)|}^{2}/2
        $
        and
        $
        \epsilon_{Bs}(t)=N_g^{-2} \sum \limits_{j,k} {
            |B_s(j\Delta_x,k\Delta_x,t)|}^2 /2 \mu_0.
        $
        The magnetic and electric fields are $B_x + iB_y$ and $E_x + iE_y$ for $s=\perp$ and $B_z$ and $E_z$
        for $s=\parallel$. The energy densities are normalised by \(\epsilon_{K0} \equiv \epsilon_{K}(t=0)\). The 
        kinetic energy density
        $
        \epsilon_{K}(t)=N_g^{-2}\Delta_x^{-3}\sum \limits_j m_{cp}
        \, c^2(\gamma_j-1).$

        The growth of $\epsilon_{B\perp}$ is exponential for \(t\omega_p<60\)  and we notice a weak 
        growth of $\epsilon_{E\parallel}$, while the other energy densities are at noise levels. This defines the 
        linear phase of the instability. The $\epsilon_{E\perp} \gg \epsilon_{E\parallel}$ at this early time, 
        because the latter is electromagnetic in the chosen geometry, while $\epsilon_{E\perp}$ contains also 
        electrostatic noise. After $t\omega_p \approx 60$ the growth of $\epsilon_{B\perp}$ slows down, while the
        other energy densities grow rapidly. Processes must be at work, which are not captured by the solution of 
        the linear dispersion relation. The $\epsilon_{B\parallel}$ is amplified here, but not in a 1D simulation 
        that evidences only a growth of $\epsilon_{E\perp}$ and of $\epsilon_{E\parallel}$\cite{stodie}. All 
        energy densities saturate at the same time at \(t\omega_p \approx 80\). Furthermore, the energy density
        \(\epsilon_{E\perp}\) is two to three orders of magnitude smaller than \(\epsilon_{B\perp}\), thus the 
        amplitude of the mean electric field is one to two orders of magnitude lower than the amplitude of the 
        mean magnetic field. Most electrons have velocities well below \(c/10\) and they are therefore affected by 
        the electric force as much as by the magnetic force. Neglecting the electric field \cite{mor} is thus not
        necessarily permitted.

        We can approximate with an exponential function the early stage of the evolution of the energy densities 
        of the perpendicular components (not shown), yielding the growth rates \(\sigma_{B\perp} \approx 0.07 
        \omega_p \), which is below the analytical value \( \sigma_{Ana} =  0.114 \omega_p\), while \(\sigma_{E\perp} 
        \approx 0.06 \omega_p\). It is the same situation as in reference \cite{stodie}: The growth rate of the 
        magnetic energy density is reduced in comparison to the analytical value. This is at least partially due 
        to the averaging over all wave numbers. The growth rate of $\epsilon_{E\perp}$ is here below that of 
        $\epsilon_{B\perp}$, while it grew twice as fast in a 1D simulation \cite{stodie}. Such a reduction of
        the growth rate of $\epsilon_{E\perp}$ as we go from a 1D to a 2D simulation is also observed for 
        the FI \cite{die09}.
        
        In Figure \ref{fig:EnDen} (b) the decrease of the thermal anisotropy is shown. With the transition to the nonlinear phase the thermal temperatures have almost equalized. Also the kinetic energies of particles moving in parallel and perpendicular direction balances (see Figure \ref{fig:EnDen} (c)).
\subsection{The power spectra of the field components}

\begin{figure}[ht!]
	  \centering
          \includegraphics[width=5cm]{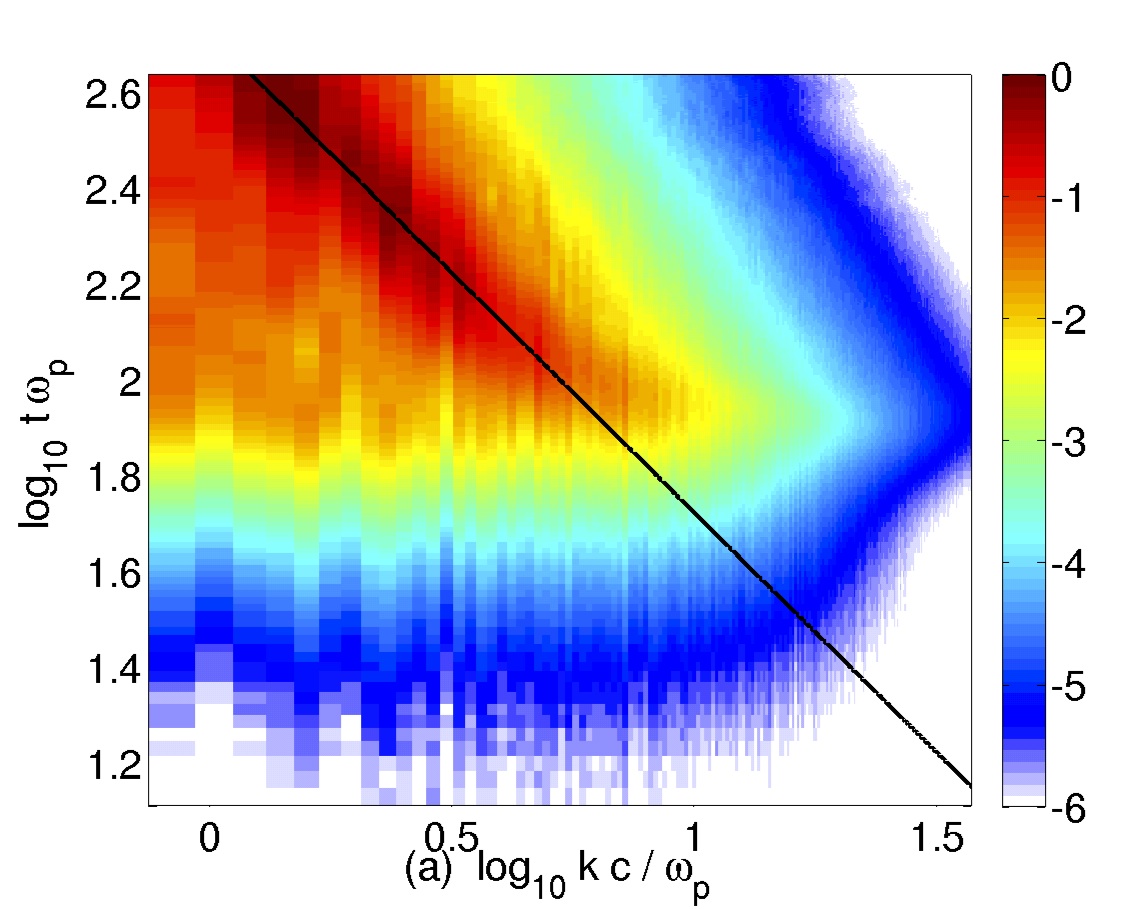}
          \includegraphics[width=5cm]{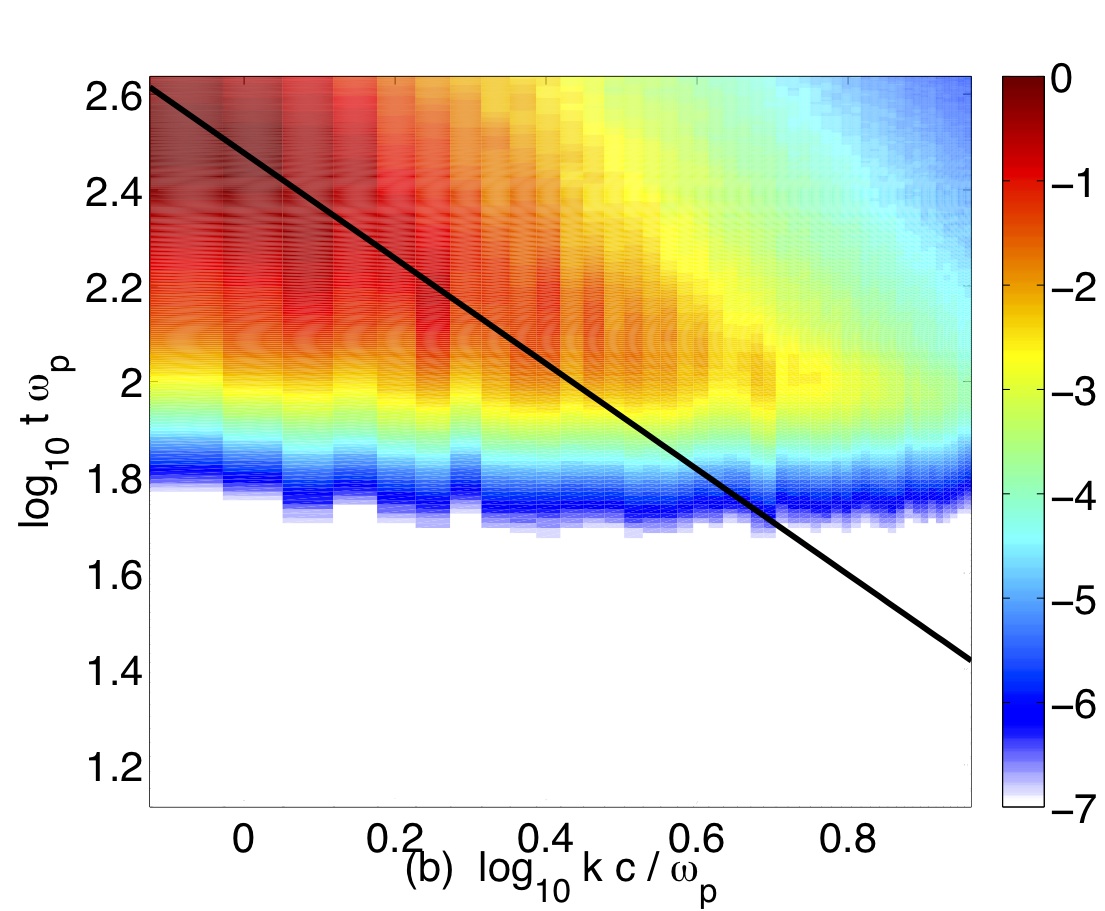}
          \includegraphics[width=5cm]{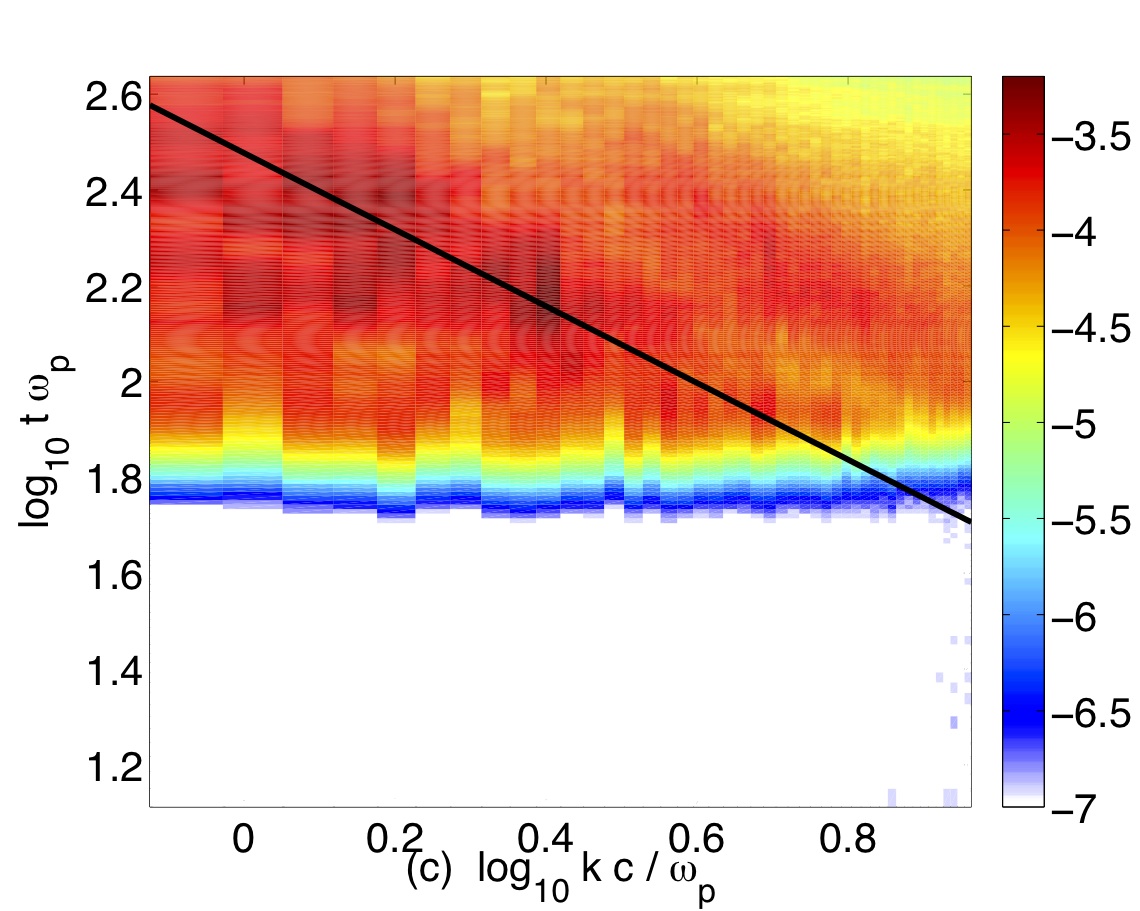}
	  \caption{(Colour online) The power spectra \(P (k,t)/P_0\) of \(cB_\perp\)
            (a), \(cB_\parallel\) (b) and \(E_\perp\) (c) as a function of the logarithmic (base 10) wave number
            and time. \(P_0\) is the maximum value of the power spectrum of \(cB_\perp\). The colourbar is the 
            10-logarithm of the normalised power. The overplotted curve \( k' \propto t^{-1}\)follows the evolution 
            of the power maximum to low \(k\).}\label{fig:Pow1}
        \end{figure}

        In what follows, we analyse the nonlinear evolution of the fields. Figures \ref{fig:Pow1}(a-c) 
        display the power spectra of \(cB_\perp\), \(cB_\parallel\) and \(E_\perp\). The power spectra as
        a function of the scalar wavenumber $k = {(k_x^2+k_y^2)}^{1/2}$ have been obtained by the integration 
        over the azimuth angle of the power spectrum in the $k_x-k_y$ plane. We have normalised the spectra 
        by \(P_0\), which is the highest value of the power spectrum \(P(k,t)\) of $B_\perp$.

        The $cB_\perp$ grows exponentially and to a significant power over a wide range of wave numbers up to 
        a $k=k_{max,Sim}$ with $\log_{10}(k_{max,Sim}c/\omega_p) \approx 1.2$. This is close to the analytical 
        value $\log_{10}(k_{max,Ana} c/ \omega_p) = \log_{10} \sqrt{A} \approx 1.3$. The $P/P_0$ gradually goes over 
        into noise at higher $k$. The magnetic field components \(c B_\perp\) and \(c B_\parallel\) grow in a 
        similar range of wave numbers, which suggests that they are coupled. This mechanism must be nonlinear, 
        because the solution of the linear dispersion relation does not predict the growth of \( c B_\parallel \). 
        The correlation of both magnetic components is reduced at later times, when the spectrum of \(c B_\perp\) 
        broadens. The wavenumber spectra of \( E_\perp \) and of \( B_\parallel \) evidence the onset of the growth 
        of structures at about the same time $t\omega_p \sim 10^2$. However, the power spectrum of the electric field
        reaches larger wave
        numbers. The electric field thus changes more rapidly in space than the magnetic one, as it would be the 
        case if it were driven for example by the magnetic pressure gradient $\propto \nabla \mathbf{B}^2$.

        Figure \ref{fig:Pow1} furthermore shows that the power of the field components during the non-linear 
        phase shifts in time to lower wave numbers $k$. This goes along with an increase of the scale size 
        (coherence length) of the magnetic field by the merging of the current structures in position space. 
        The merging requires two spatial dimensions orthogonal to the parallel axis \cite{mor}. In a 1D 
        simulation the merging of the current filament eventually stalls \cite{mor,stodie,pal2009}. The wave 
        number associated with the maximum power of $cB_\perp$ can be approximated by the curve $k' \propto 
        t^{-1}$. However, the power spectra are broadband and the sampling time too short to determine 
        accurately this dependence. 
        \begin{figure}[ht!]
          \centering
          \setlength{\unitlength}{0.001\textwidth}
          \begin{picture}(1100,450)(-260,0)
            \includegraphics[width=7cm]{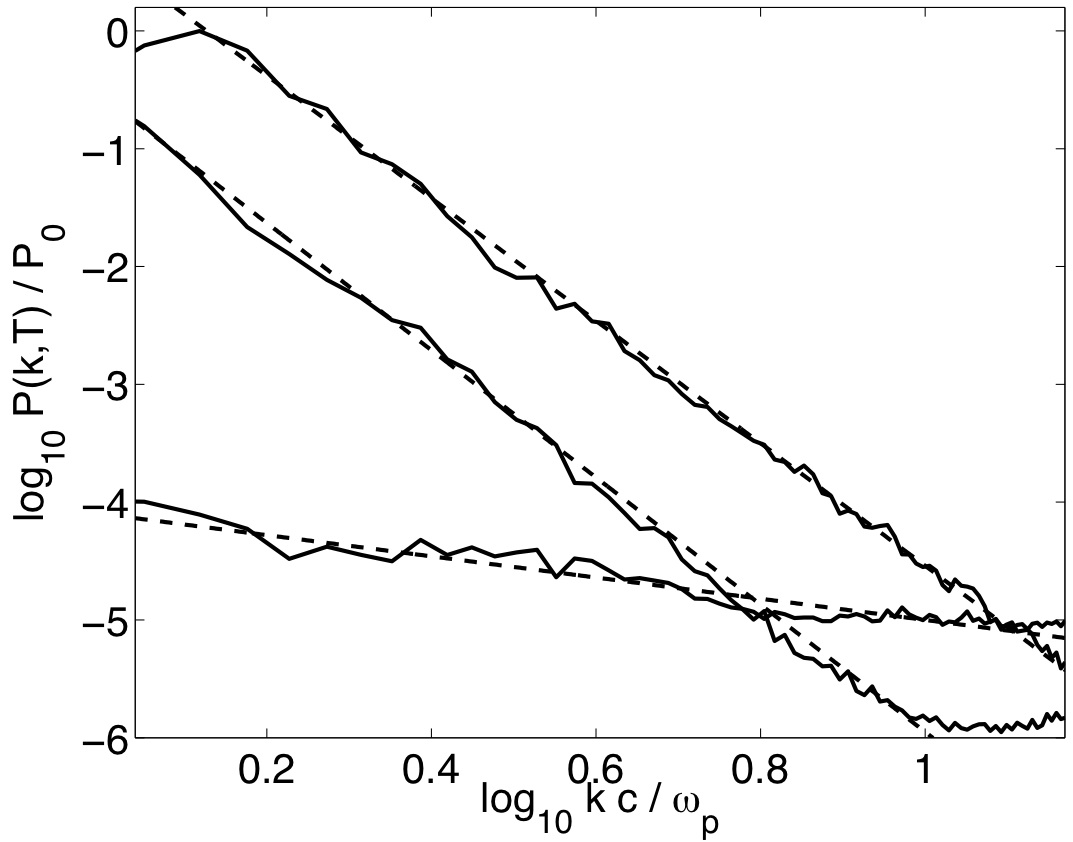}
            \put(-180,220){\small \(cB_\perp\)}
            \put(-320,205){\small \(cB_\parallel\)}
            \put(-300,95){\small \(E_\perp\)}
          \end{picture}
          \caption{10-logarithmic plot of the power spectrum at \(t
            \omega_p=434\). The dashed lines are the fits to the
            simulation data.}\label{fig:Pvonk}
        \end{figure}     

        Figure \ref{fig:Pvonk} is a double logarithmic plot of the power spectra of the field 
        components $c B_\perp$, $cB_\parallel$ and $E_\perp$ against the normalised wave number 
        at the end of the simulation. Only the range up to $\log_{10} (kc/\omega_p)=1.15$ is 
        shown here as the noise dominates at large values. The power spectra have a power-law
        behaviour for almost the whole range. The indices are $-5.2$ for $cB_\perp$, $-5.4$ for 
        $cB_\parallel$ and $-0.9$ for $E_\perp$. The power spectra of $B_{\perp}$ and $B_\parallel$ 
        are practically identical, except for a scaling factor $\approx 10$. This is further evidence
        for a nonlinear coupling between both components. In contrast, the power spectrum of the 
        electric field is flat compared to the magnetic components. The amplitudes for $cB_\perp, 
        cB_\parallel$ and $E_\perp$ are proportional to a force. The spectral distributions reveal a 
        trend, namely that the electric forces can become more important than the magnetic ones, if 
        the dynamics of non-relativistic electrons on small scales (large $k$) is considered.
        
\subsection{The saturation mechanism}
        The field data are now analyzed in order to give information about the saturation mechanism and about
        how the driving field $B_\perp$ couples to $E_\perp$ and to $B_\parallel$.  We consider for this purpose the equation 
        \begin{equation}
          \partial_t (n \mathbf v) + \nabla \cdot (n \mathbf v \mathbf
          v ) = -\frac{1}{m_e} \nabla \cdot \hat P_T - \frac{e n}{m}
          (\mathbf E + \mathbf v \times \mathbf B),
        \end{equation}
        with which we can investigate the interplay of the field components. It describes the spatio-temporal 
        evolution of a single-species fluid consisting of electrons with particle density \(n\), mean velocity 
        \(\mathbf v\) and thermal pressure tensor \(\hat P_T\). With Ampere's law \(\mathbf J = \frac{1}{\mu_0} 
        \nabla \times \mathbf B - \epsilon_0 \frac{\partial \mathbf E}{\partial t}\), \(\mathbf J = -e n \mathbf v\) 
        and \((\nabla \times \mathbf B ) \times \mathbf B= - \left[ \nabla \mathbf
          B^2 / 2 - \nabla \cdot (\mathbf B \mathbf B) \right]\) the
        equation can be rewritten as
        \begin{equation}
          \partial_t (n \mathbf v) + \nabla \cdot (n \mathbf v \mathbf
          v ) = -\frac{1}{m_e} \nabla \cdot \hat P_T - \frac{e n}{m}
          \mathbf E - \frac{1}{m } \nabla \cdot \hat \sigma -
          \frac{1}{m} \nabla P_B + \frac{\epsilon_0
          }{m} \mathbf B \times \frac{\partial \mathbf E}{\partial t}
          \label{forceEQ}
        \end{equation}
        where \(\hat \sigma= \mathbf B \mathbf B / \mu_0\) and \(P_B= \mathbf B^2/2 \mu_0\)
        describe the magnetic stress tensor and pressure.

        If we consider a configuration where the gradients \(\partial_y\) and \(\partial_z\) are not resolved, 
        the term \(-\nabla \cdot \hat \sigma / m\) due to the magnetic tension vanishes. The development of 
        the FI or the TAWI imply, that the term $m^{-1} \nabla P_B$ on the right hand side of Eq. \ref{forceEQ} 
        becomes important. The PIC simulations in \cite{stodie,die09} demonstrate that an electric field grows 
        with an amplitude that cancels the term $m^{-1} \nabla P_B$ by the second term $en\mathbf{E}/m$ on the 
        right hand side of Eq. \ref{forceEQ}, provided that immobile ions are considered. The amplitude of the 
        electric field along the 1D simulation box is then given by
        \begin{equation}
          E_x = -\frac{1}{2 e n_e \mu_0} \, \frac{d}{dx} B^2 = -\frac{1}{e n_e} \frac{d}{dx} P_B
        \end{equation}
        As a result, the electric field energy density grows twice as fast as the magnetic one. The TAWI and FI 
        saturate, when the combined electric and magnetic force prevents a further spatial re-arrangement of the 
        current by confining the electrons in space. 
 
        The perpendicular gradients are resolved (\(\partial_x \not= 0\), \(\partial_y \not= 0\)) in our 2D PIC
        simulation and \(\nabla \cdot \hat \sigma\) no longer vanishes. If we neglect the contribution of the 
        thermal pressure gradient and the displacement current (term 1 and 5 on the right hand side in Eq. \ref{forceEQ}), then the
        right hand side of Eq. \ref{forceEQ} vanishes, provided that
        \begin{equation}
        \mathbf{E}_\perp = \frac{1}{en} \nabla_{xy} \cdot \hat{\sigma} - \frac{1}{en} \nabla_{xy} P_B, 
        \label{NLTerms} 
        \end{equation}
        where $\nabla_{xy}$ is the gradient in the x-y plane. We will examine here the magnitude of the three terms 
        in Eq. \ref{NLTerms} using the PIC simulation data.
        
        Movie 1 shows the development of the individual contributions: The upper left panel shows the magnetic 
        field \(|\mathbf B(x,y)|\) normalised by $\omega_p m /e$. The middle panel shows the development of the 
        nonlinear terms in Eq. \ref{NLTerms} in the same normalisation $\omega_p m / ce$ as the electric field, 
        i.\,e.\ the superposition of the magnetic pressure gradient \(\nabla P_B\) and the divergence of the 
        magnetic stress tensor \(\nabla \cdot \hat \sigma\). The bottom panel is the normalised electric field 
        component \(E_\perp(x,y)\). Besides the formation of larger structures in the magnetic field \(|\mathbf 
        B(x,y)|\), the connection of the magnetic tension and pressure gradient force with the perpendicular 
        electric field is obvious at various locations in Movie 1. For \(t \omega_p > 50\) structures in the 
        electric field are contrasted with the noise background and the strongest ones correspond to the two 
        non-linear terms we examine here.
        \begin{figure}[ht!]
          \centering
          \includegraphics[width=0.49\textwidth]{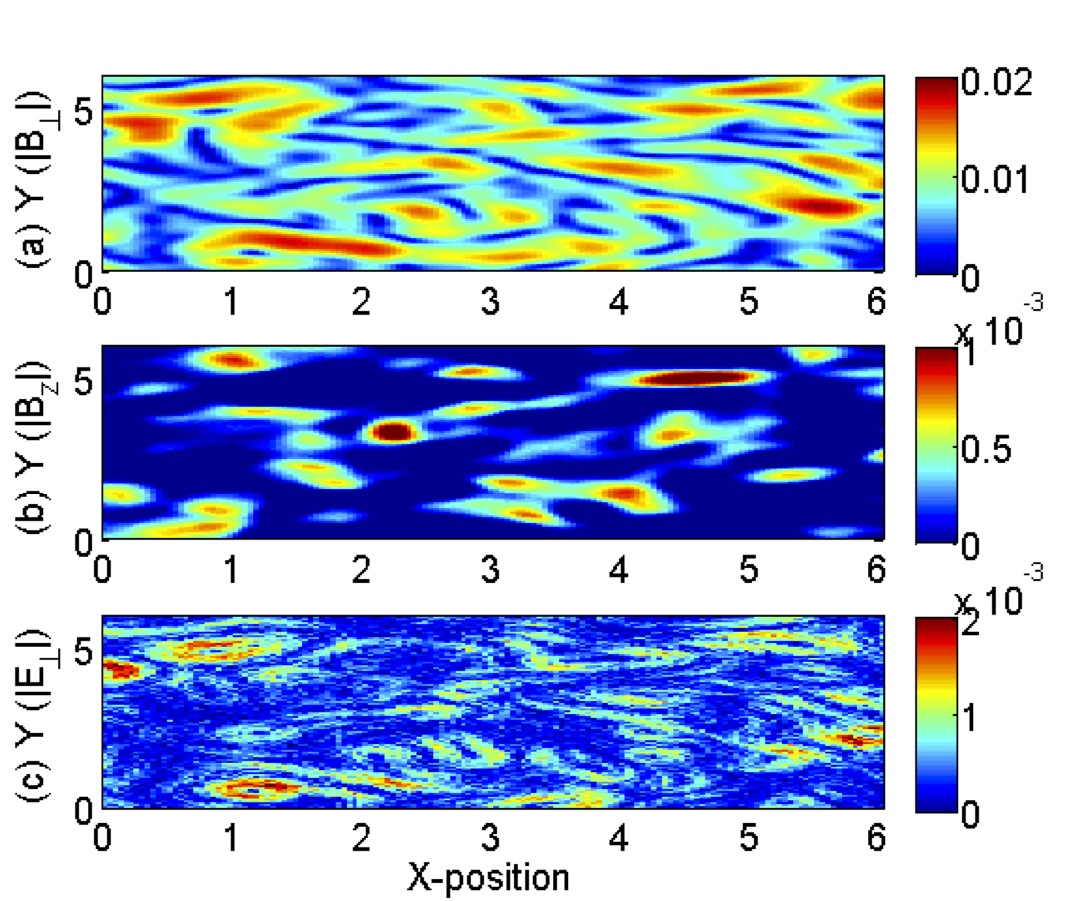}
          \includegraphics[width=0.49\textwidth]{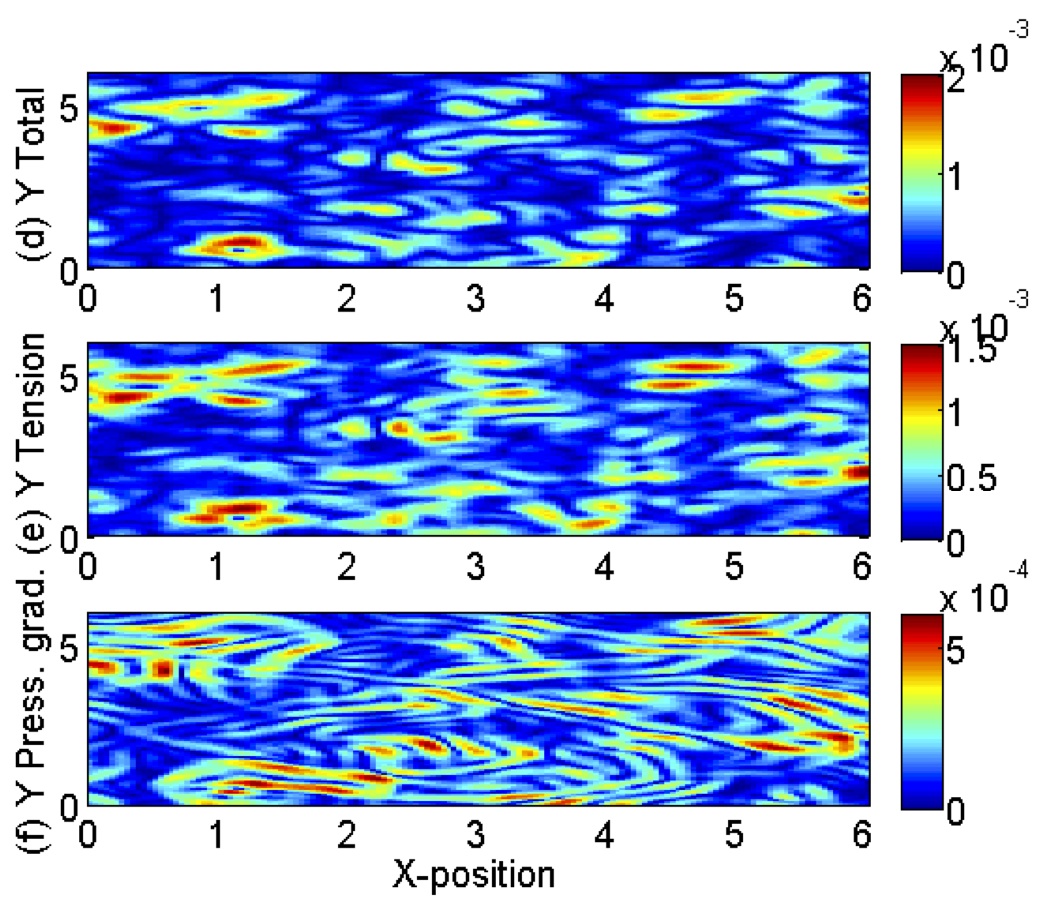}
          \caption{(Colour online) The relevant field and force components at $\omega_p t=75$ in a subinterval of the box: 
          (a) \(|B_\perp| = |B_x + i B_y|\), (b) the parallel magnetic component $|B_\parallel|=
          |B_z|$, (c) the \( |E_\perp|= |E_x +iE_y| \). The force \( |\nabla \cdot \hat \sigma + \nabla P_B |\)
          (d) and the individual components (e) \( | \nabla \cdot \hat \sigma | \) and (f) 
          \( | \nabla P_B |\). Values are normalized and can be compared directly.}\label{fig:Terms1}
        \end{figure} 
        Figure \ref{fig:Terms1} displays the most relevant field and force components at the time $\omega_p t = 75$, 
        when the TAWI has just saturated in Fig. \ref{fig:EnDen}. The dominant field component is clearly $|B_\perp|$, 
        as expected from the TAWI and from the energy density diagram. The peak amplitude $\approx 0.02$ is compatible 
        with a saturation by the magnetic trapping mechanism. The magnetic bounce frequency in the normalized units 
        $\tilde{k}=kc/\omega_p$, $\tilde{v}=v/c$ and $\tilde{B}_\perp = m B_\perp / e$ displayed in Fig. 
        \ref{fig:Terms1} is $(\omega_B / \omega_p) = {|\tilde{k}\tilde{v}\tilde{B}_\perp|}^{1/2}$. We rearrange the
        equation and get $\tilde{v} = {(\omega_B / \omega_p)}^{2}/(|\tilde{B}_\perp| \tilde{k})$. Equating the 
        $\omega_B$ with the typical growth rate from Fig. \ref{Plot1} gives $\omega_B / \omega_p = 0.05$ and we further
        assume that $\tilde{k}=2.5$ and $|B_\perp| = 0.02$. We get the reasonable rough estimate $\tilde{v} \approx 0.05$,
        which is a few times $v_{th\parallel} /c \approx 0.01$.  
        
        The correspondence between the total force $|\nabla \cdot \hat \sigma + \nabla P_B |$ and $|E_\perp|$ is 
        apparent. The amplitudes are very similar and at least the strong structures in Figs. \ref{fig:Terms1}(c) 
        and (d) agree well, e.\,g.\ at \(x \omega_p /c = 1\) and \(y \omega_p / c = 0.5\) or at \(x \omega_p /c = 0.5\) 
        and \(y \omega_p / c = 4.5 \). We find that $|E_\perp| \approx |B_\perp|/10$ in various locations and the 
        electric force on an electron moving with the speed $c/10$ will equal the magnetic one. We decompose the total
        force into its constituents in Fig. \ref{fig:Terms1}(e,f). The major force contribution arises from the 
        magnetic tension force. However, the omission of the magnetic pressure gradient in Fig. \ref{fig:Terms1}(e) 
        alters the force distribution and results in clear differences with regard to $|E_\perp|$. It is thus obvious 
        that both components provide important contributions to the total electric field. No clear connection between
        $B_\parallel$ and the other components is visible in Fig. \ref{fig:Terms1}.
        
\begin{figure}[ht!]
          \centering   
          \includegraphics[width=7cm]{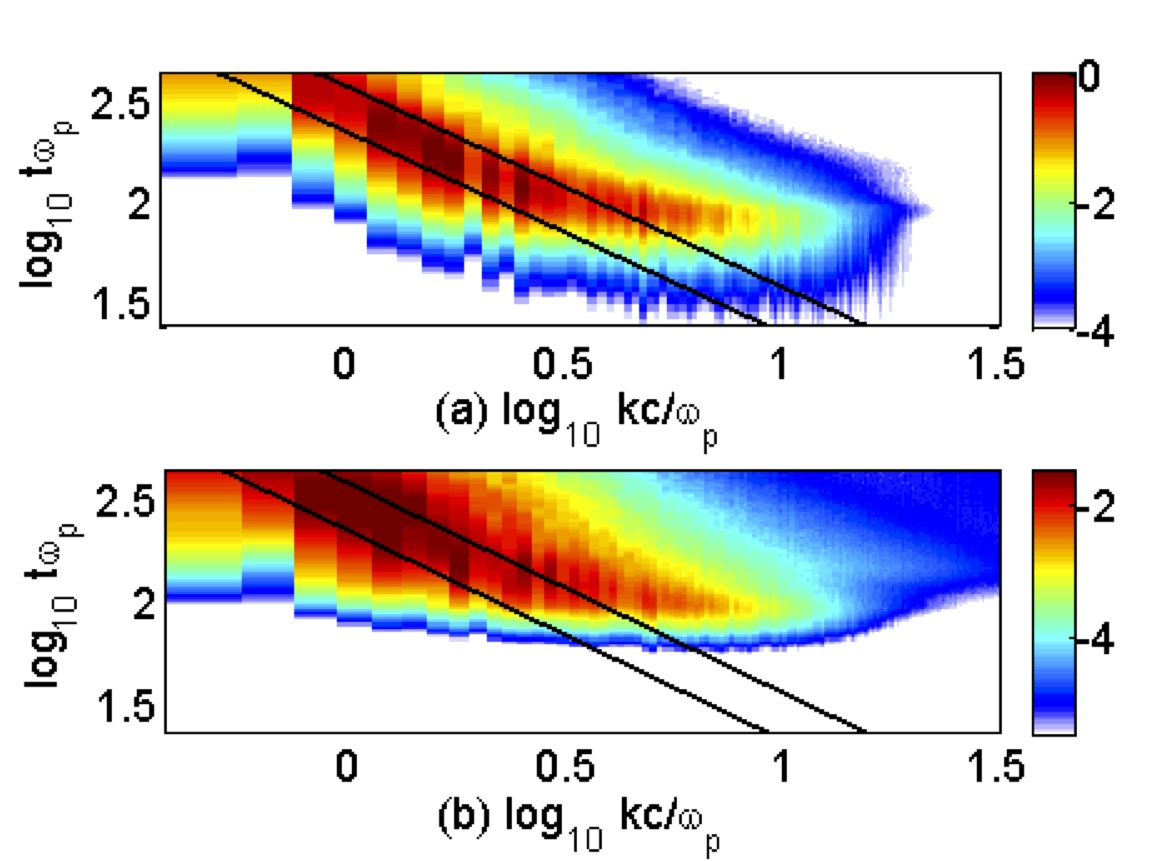}
          \caption{(Colour online) The power spectra $P(k,t)$ of the current components \(J_\parallel\) (a) and \(J_\perp\) (b).
          Both spectra are normalized to the maximum power of \( J_\parallel\). The wavenumber, the time and the 
          colour scale are 10-logarithmic. Overplotted are two curves $k\propto t^{-1}$, which confine the
          wavenumber interval with the maximum power. The curves are the same in both panels.}\label{fig:powSpecJ}
        \end{figure}

        The currents can provide further information about the coupling between $B_\parallel$ and the other field
        components. We subdivide for this purpose the current into a $J_\parallel = J_z$ and $J_\perp = J_x +
        iJ_y$. The power spectrum of both components is calculated by a 2D Fourier transform over the $x-y$ plane
        followed by the azimuthal integration and a normalization by the maximum power of $J_\parallel$. This provides 
        us in analogy to the field distribution Fig. \ref{fig:Pow1} with the time-dependent power spectra of 
        $J_\parallel$ and $J_\perp$, which we display in Fig. \ref{fig:powSpecJ}.
        
        The expectation that $J_\parallel \gg J_\perp$ due to $|B_\perp| \gg |B_\parallel|$ is confirmed. Both 
        current distributions reveal that the wavenumber intervals, in which the power spectra of the currents 
        peak, can be confined by two curves that follow $k\propto t^{-1}$. The exact dependence 
        of the wavenumbers, at which the power spectra of the currents peak is, however, not a power of $t$. 
        The latter would correspond to a straight line in the double logarithmic diagram. Both current distributions 
        rapidly expand to lower $k$ after $t\omega_p \approx 125$. This broadening of the current distribution is 
        probably connected to that of the fields at late times in Fig. \ref{fig:Pow1}. A finite box effect may be
        responsible for this sudden broadening, because the largest observed wavelengths are not longer small 
        compared to the box size. We discuss one such finite box effect below. However, this broadening sets in well 
        after the saturation time $t\omega_p \approx 75$ and these two processes are not connected. The signal-to-noise
        ratio of the currents is sufficient to reveal the cut-off of both spectra at $log_{10}(kc/\omega_p) =
        \log_{10}\sqrt{A}$, which is in line with Fig. \ref{Plot1}.

        The overplotted curves $k\propto t^{-1}$ show that the wavenumber intervals, in which the power spectra
        of both current components peak, are practically identical. This confirms their connection. The abrupt 
        growth of the power spectrum of $J_\perp$ at $t\omega_p \approx 70$ furthermore suggests that it is caused 
        by a non-linear process. The immobile ions imply, that the $J_\perp$ is caused by the bulk motion of the 
        electrons in the simulation plane. The electrons can be accelerated either by $E_\perp$ or by their 
        deflection by $B_\perp$ from the parallel direction into the perpendicular plane. The electromagnetic 
        fields that accelerate the electrons in the perpendicular plane are tied to the spatially non-uniform 
        $J_\parallel$, which is driven by TAWI. The fields' extent is comparable to the spatial size of the current
        filaments. It is thus not surprising that the current structures in $J_\perp$ have a typical size that is 
        comparable to that of the structures in $J_\parallel$. 
        
        The time evolution of the current distributions $J_\parallel (x,y)$ (upper panel) and $J_\perp (x,y)$ (lower 
        panel) are shown in the movie 2 for a subsection of the simulation box. The merging of filaments in $J_\parallel$
        can be observed. Some merge to chains. If stable chains form, that spread across the entire simulation box with 
        its periodic boundary conditions, then this would result in finite box effects. The formation of such 
        long chains of filaments may cause the sudden spread of the currents to low $k$ in Fig. \ref{fig:powSpecJ} 
        and in the electromagnetic fields in Fig. \ref{fig:Terms1}. The current structures in $J_\perp$ are driven by 
        the fields and the current is dissipated away, resulting in their limited lifetime. The structures in $J_\perp$,
        e.g. the current eddies in the lower panel of movie 2, give rise to the $B_\parallel$ in Fig. \ref{fig:Terms1}.       

\subsection{The velocity distribution}
        
\begin{figure}[ht!]
	  \centering
          \includegraphics[width=7cm]{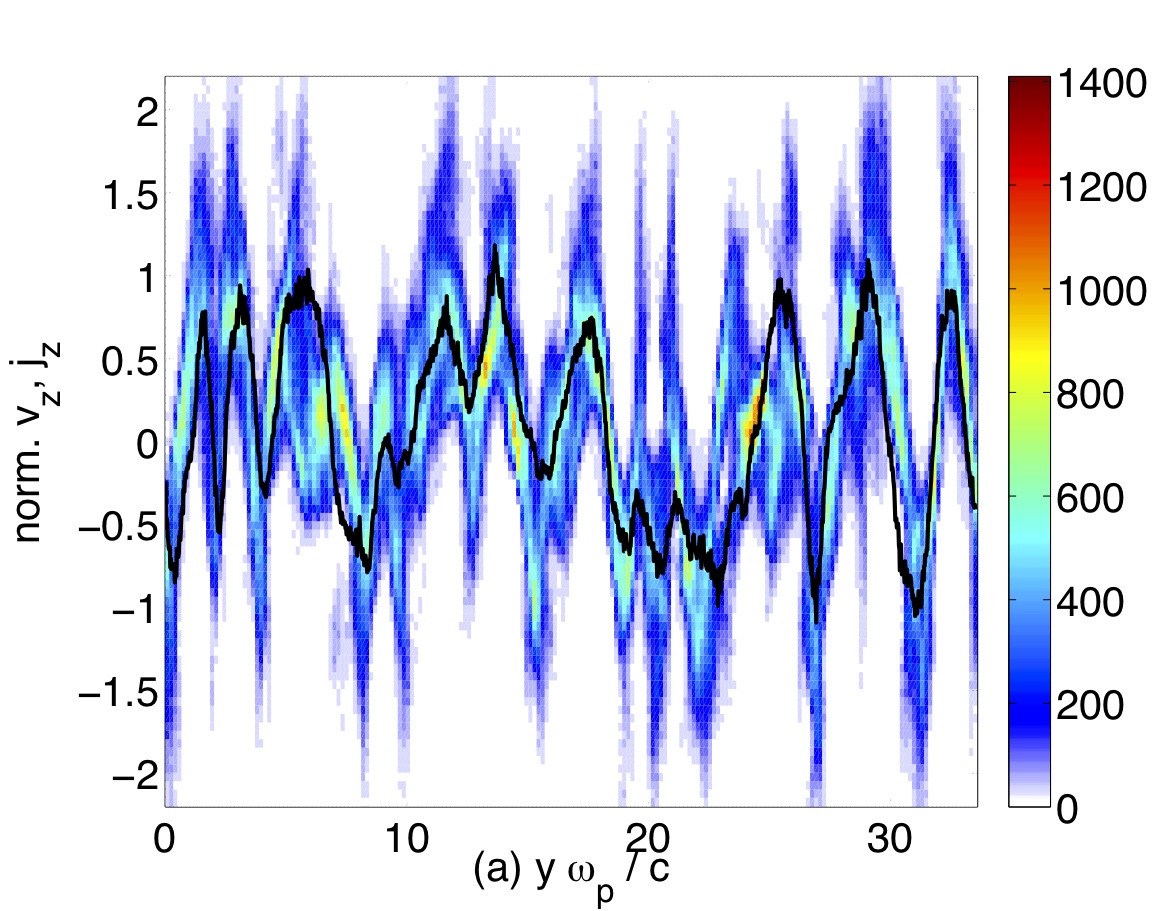}
          \hspace{3pt}
          \includegraphics[width=7cm]{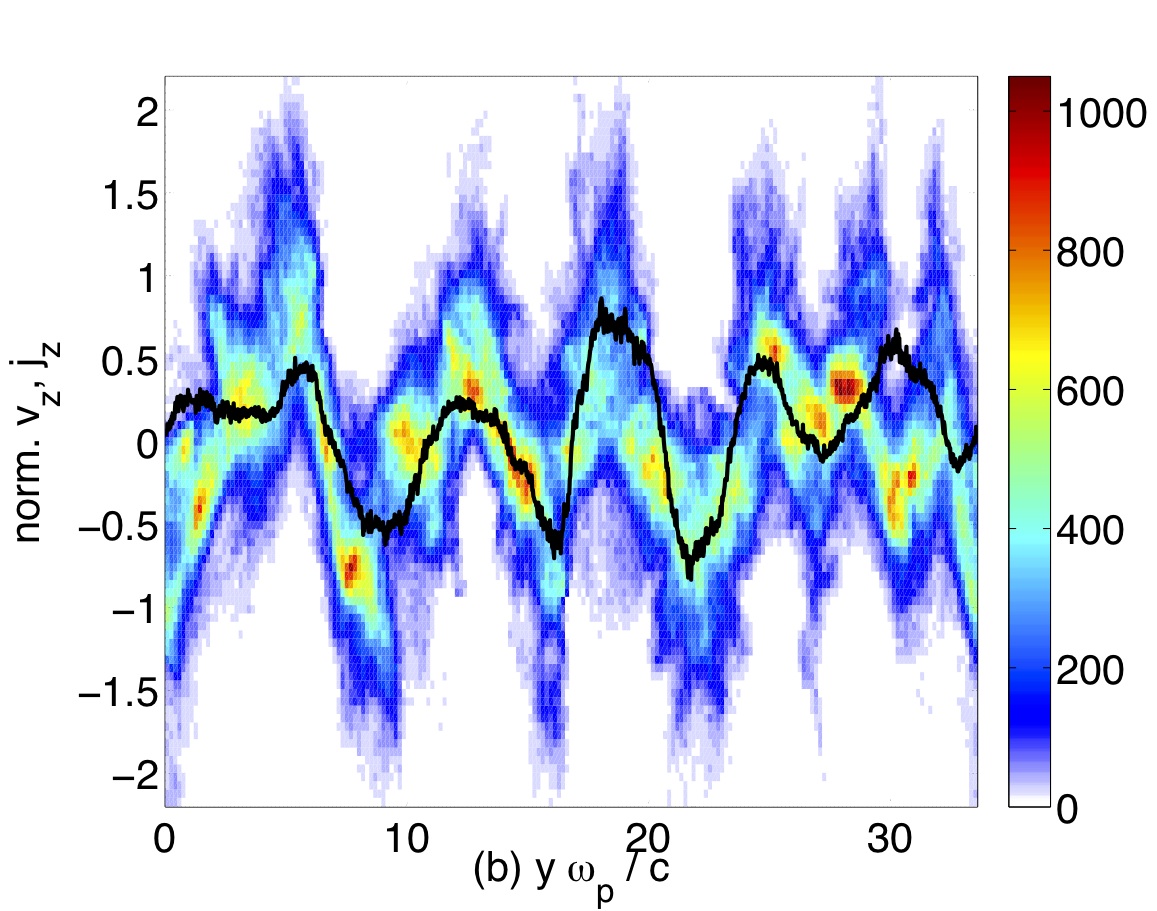}
          \vspace{4pt}
          \includegraphics[width=7cm]{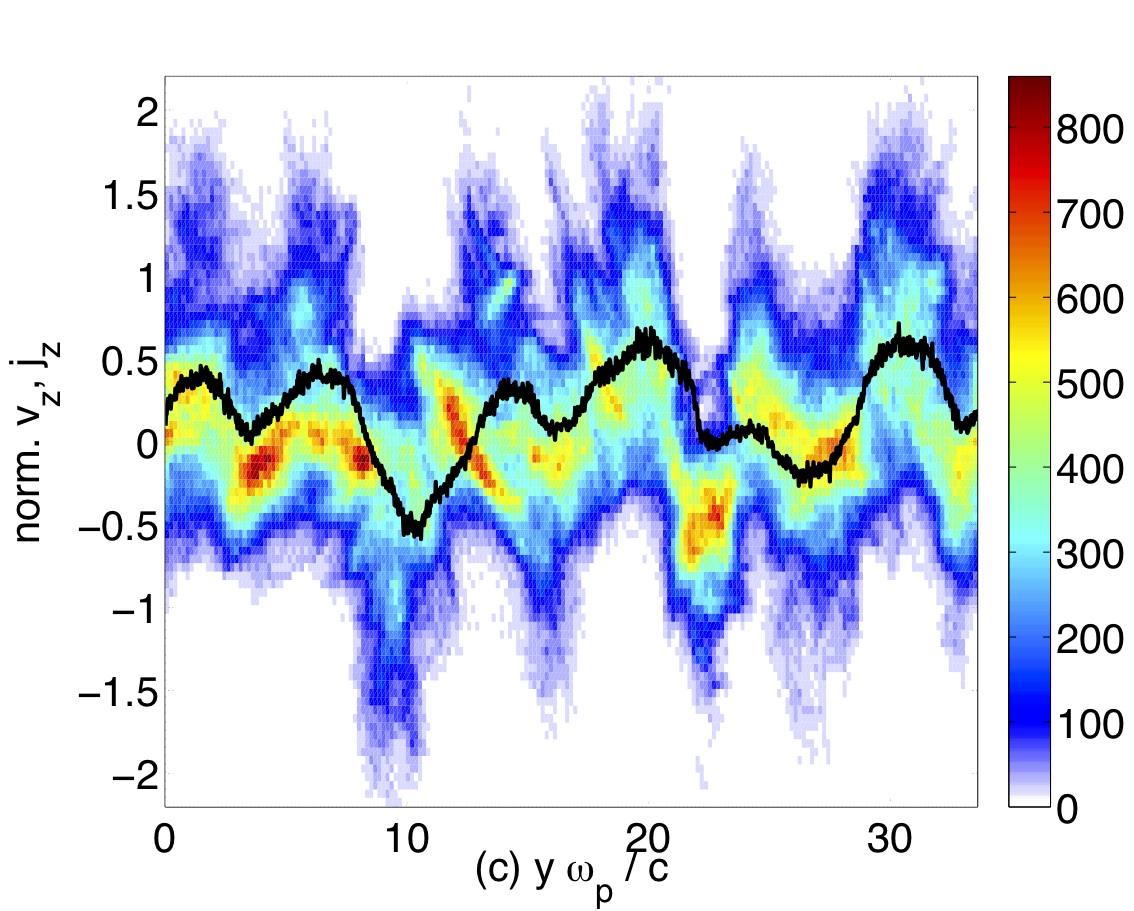}
          \hspace{3pt}
          \includegraphics[width=7cm]{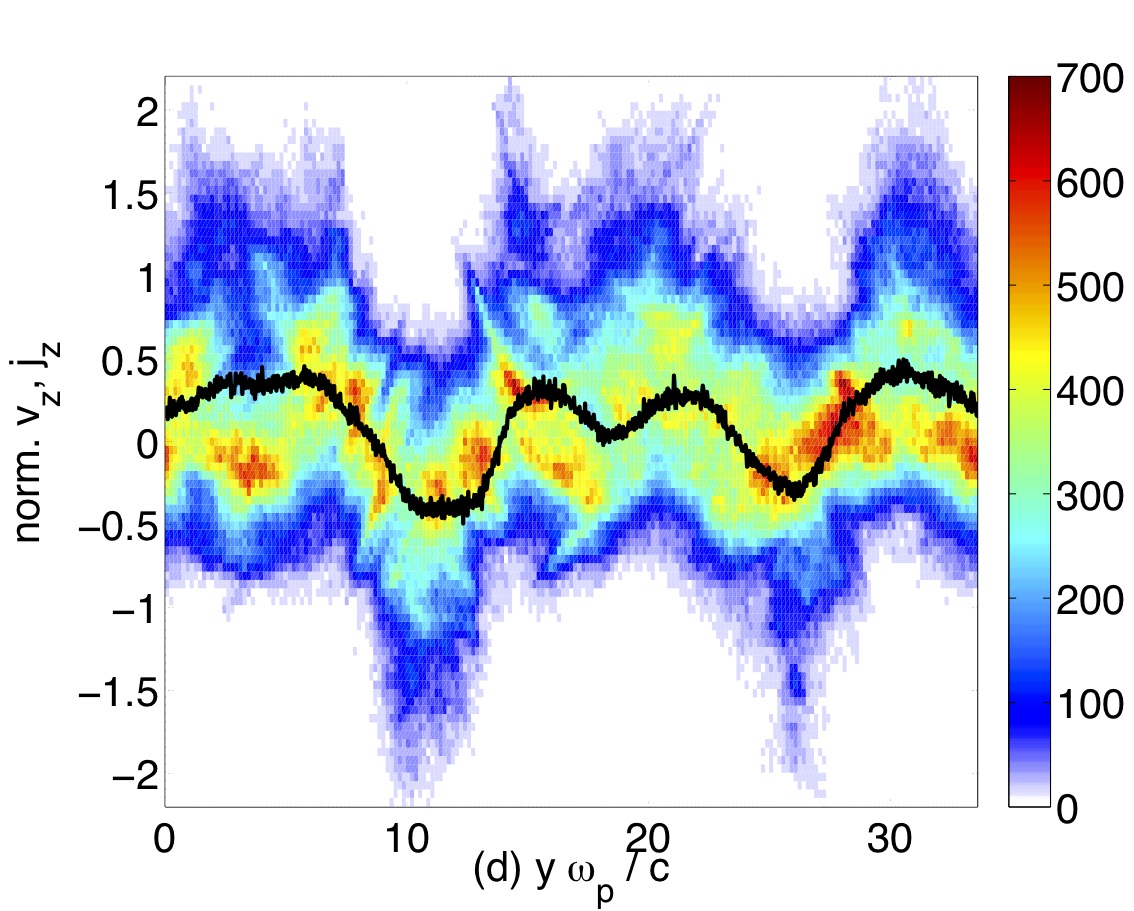}
	  \caption{(Colour online) The colour plot shows the normalised velocity
            distribution $v_z(y) / v_{th\parallel} $ for $x \omega_p /
            c = 16.8$ at the simulation times $t \omega_p = 118$ (a),
            198 (b), 278 (c) and 358 (d). Overplotted is the
            normalised current density $j_z(y) n_e /q_{cp}
            c$.}\label{fig:VelCurr}
        \end{figure}

        The strong $J_\parallel$ should visibly modulate the phase space distribution $f(x,y,v_z)$. It is, however, 
        unclear if the phase space density remains compact or if beams form, i.e. distinct electron distributions for a
        given position along $x$ and $y$. Figure \ref{fig:VelCurr} displays the phase space distribution as a function of
        $v_\parallel (y)$ for a fixed $x \omega_p / c = 16.8$. This phase space diagram is shown for the four simulation
        times $t \omega_p =$ 118, 198, 278 and 358. The velocity distribution, which has initially been spatially uniform,
        has been transformed into one, for which the mean speed varies as a function of the position. This velocity
        rearrangement has been accompanied by an energy transfer into the magnetic field.
        
        At $t \omega_p = 118$, just after the TAWI has saturated, the velocity distribution oscillates in space with 
        a high frequency, which corresponds to high $k$ in Fig. \ref{fig:powSpecJ}. The spatial
        oscillation frequency decreases with time, while the thermal spread of the electron beam increases. The 2D
        distribution is scanned along the $x$-direction shown in Movie 3 at \(t\omega_p = 358\), showing that the
        velocity oscillations along the slice $x \omega_p / c = 16.8$ in Fig. \ref{fig:VelCurr} are representative 
        for those in the simulation box. The actual velocity spread, within which we find significant numbers of 
        electrons (note the linear density scale) is the same during the non-linear phase with (\(|v_z|/ 
        v_{th\parallel} \lesssim 1.5\)). Faster particles do exist. Initially the Maxwellian along the parallel 
        direction could be populated with the given statistical plasma representation of 160 particles per cell up 
        to $|v_{max}|\approx 2.5 v_{th\parallel}$. The low number density of the fast particles implies, however, that 
        they do not contribute much to $J_\parallel$. The simulation furthermore evidences, that no electrons are
        accelerated to high speeds during the non-linear evolution of the TAWI despite the large initial value of $A$. 
 
        The increase with time of the wave length of the spatial oscillations in Fig. \ref{fig:VelCurr} is related to 
        a merging of the filaments in the $J_\parallel$ distribution (Movie 2), which could not be observed to this 
        extent in the 1D simulation of our previous work \cite{stodie}. In the latter work, the particles have instead 
        been pushed together by the magnetic pressure gradient, yielding a layered structure in the velocity distributions.
        In our 2D simulation the filament merging destroys this effect. In agreement with these results, the structure
        becomes more diffuse and the normalised current density $j_z n_e / q_{cp} c = J_z /q_{cp} c$, overplotted in 
        Fig.\ \ref{fig:VelCurr}, with the current follows the structure of the velocity distribution.

\section{Discussion}\label{discussion}
        The thermal anisotropy-driven Weibel instability (TAWI), which is an important seed mechanism for magnetic
        fields in astrophysical environments and in laser-generated plasmas \cite{Bell,sch2,kar}, has been investigated 
        here with the help of a particle-in-cell (PIC) simulation for mobile electrons and for immobile ions. The 
        plasma has initially been spatially uniform and we have set the initial electric and magnetic fields to zero. 
        The thermal speed $v_{th\parallel}$ of the electrons along 
        one direction has been 20 times larger than the equivalent $v_{th\perp}$ in the perpendicular plane. The 
        anisotropy parameter $A=v_{th\parallel}^2/v_{th\perp}^2 -1 = 399$ is thus very large. It allows us to examine 
        the field growth and the electron thermalization under extreme conditions. It also provides us with a good 
        signal-to-noise ratio of the electromagnetic fields in simulation. Thus, the results are valid for high anisotropies only. We plan to discuss the low anisotropy case in future work.
        
        The main purpose of our work has been to determine the source mechanism of the electric field during the
        non-linear stage of the instability in more than one dimension. This electric field has recently received
        attention \cite{kaa09,pal2009}, because the electric force is not small compared to the magnetic one. Both 
        field components are thus important for the saturation of the TAWI, while typically only the magnetic field
        is considered \cite{mor,dav02}. We have previously determined that the electric field is driven by the magnetic
        pressure gradient force, if the wave spectrum is limited to one dimension \cite{stodie}. This force component 
        is dominant for the filamentation instability in 1D and 2D simulations \cite{die09}, but it was not clear if 
        this finding holds also for the TAWI.  
    
        We summarize our results: We have compared the interval of unstable wave numbers in the simulation with the 
        corresponding solution of the linear dispersion relation. Both agree prior to the saturation of the TAWI.
        Thereafter, the merging of the current filaments implies that the peak in the power spectrum of the current 
        moves to lower wave numbers. This characteristic wave number decreases approximately linearly with an increasing
        time. The scale size of the filaments and the coherence length of the magnetic and the electric fields thus
        increases
        approximately linearly in time in the position space, until a sudden broadening sets in, which we have attributed
        to finite box effects. This evolution of the filament size contrasts the results of 1D simulations, where mergers
        are possible only until the magnetic field becomes strong enough to keep the filaments with oppositely directed
        currents separated \cite{mor,stodie,kaa09}. The repelling filaments can go around each other in a 2D plane and
        they continue to merge with other filaments, which have the same direction of the current vector. 
        
        We have found that in a 2D simulation the magnetic tension force becomes stronger than the magnetic pressure
        gradient force, which clearly distinguishes this non-linear system from that driven by the beam filamentation
        instability. Both forces arise from the interaction of the net current driven by the TAWI with the magnetic
        field it generates. The electric field strength and distribution in the simulation plane resemble that of the 
        force distribution obtained from the summation of the magnetic tension and pressure gradient force, at least 
        in what concerns the strongest electric field structures. This electric field reaches an amplitude that makes 
        it equally important for the dynamics of the slow electrons as the magnetic field. Its energy density remains,
        however, well below the magnetic one, which is in agreement with previous simulations. We find that the estimate
        for the magnetic field amplitude, which lets the TAWI saturate, that is based on the magnetic trapping mechanism
        is still a reasonable approximation for that observed in our simulation. This has also been reported for 1D
        simulations of the non-relativistic TAWI \cite{kaa09} and for the filamentation instability \cite{die09}. 
        
        The electromagnetic fields generated by the TAWI drive electron currents in the simulation plane. These
        currents result in the growth of a magnetic field component, which is orthogonal to the simulation plane 
        and parallel to the direction, along which the electrons are hottest. The growth of such a field component is 
        not predicted by the solution of the linear dispersion relation and it is thus a purely non-linear process. This
        magnetic field component does also not grow in 1D PIC simulations \cite{stodie}. Its amplitude is lower than 
        that in the simulation plane but it is large enough to result in complicated 3D magnetic patterns in a 3D
        simulation \cite{rom}.
                 
        The ratio of the energy density of the two magnetic field components in our simulation plane and the total
        electron thermal energy density exceeds the expected limiting value of 1/12 \cite{lem1}. It remains, however, 
        just below twice that value. Such a peak energy density 
        of the magnetic field in the simulation plane is reasonable, if the limit 1/12 applies to each magnetic degree 
        of freedom. One magnetic component is considered in Ref. \cite{lem1}, while two components grow in our
        simulation plane.

{\bf Acknowledgements:} This work was partially supported by the
Deutsche Forschungsgemeinschaft through grant Schl 201/21-1, the
Research Department Plasmas with Complex Interactions at
Ruhr-University Bochum and by Vetenskapsr\aa det. We thank the HPC2N
supercomputer centre for the computer time and support.


\section*{References}


\begin{thebibliography}{99}

\bibitem{Bell} Bell A R and Lucek S G 2001 {\it Mon. Not. R. Astron. Soc.} {\bf 321} 433

\bibitem{sch2} Schlickeiser R and Shukla P K 2003 {\it Astrophys. J.} {\bf 599} L57

%\bibitem{sch1} Schlickeiser R 2004 {\it Phys. Plasmas} {\bf 11} 5532

%\bibitem{schr} Schaefer-Rolffs U and Schlickeiser R 2006 {\it Phys. Plasmas} {\bf 13} 012107

%\bibitem{aro} Arons J 2008 {\it AIPC} {\bf 983} 200

%\bibitem{puk} Pukhov A 2001 {\it Phys. Rev. Lett.} {\bf 86} 3562

\bibitem{yoo} Yoon P H and Davidson R C 1987 {\it Phys. Rev. A} {\bf 35} 2718

\bibitem{tau} Tautz R C and Schlickeiser R 2006 {\it Phys. Plasmas} {\bf 13} 062901

\bibitem{ach} Achterberg A and Wiersma J 2007 {\it Astrom. Astrophys.} {\bf 475} 1

\bibitem{pet} P\'etri J and Kirk J G 2007 {\it Plasma Phys. Control. Fusion} {\bf 49} 1885

\bibitem{sil1} Silva L O, Fonseca R A, Tonge J W, Dawson J M, Mori W B and Medvedev M V 2003 {\it Astrophys. J.} 
{\bf 596} L121

\bibitem{sto1} Stockem A, Dieckmann M E and Schlickeiser R 2008 {\it Plasma Phys. Control. Fusion} {\bf 50} 025002

\bibitem{med} Medvedev M V and Loeb A 1999 {\it Astrophys. J.} {\bf 526} 697

\bibitem{kar} Karmakar A, Kumar N and Pukhov A 2009 {\it Phys. Rev. E} {\bf 80} 016401

\bibitem{mor} Morse R L and Nielsen C W 1971 {\it Phys. Fluids} {\bf 14} 830

\bibitem{lem1} Lemons D S, Winske D and Gary S P 1979 {\it J. Plasma Phys.} {\bf 21} 287

\bibitem{lem2} Lemons D S and Winske D 1980 {\it J. Plasma Phys.} {\bf 23} 283

\bibitem{boro} Borodachev L V and Kolomiets D O 2010 {\it J. Plasma Phys.}, doi:10.1017/S0022377810000188, in press

\bibitem{kaa09} Kaang H H, Ryu C M and Yoon P H 2009 {\it Phys. Plasmas} {\bf 16} 082103

\bibitem{pal2009} Palodhi L, Califano F and Pegoraro F 2009 {\it Plasma Phys. Controll. Fusion} {\bf 51}, 125006

\bibitem{stodie} Stockem A, Dieckmann M E and Schlickeiser R 2009 {\it Plasma Phys. Control. Fusion} {\bf 51} 075014

\bibitem{die09} Dieckmann M E 2009 {\it Plasma Phys. Control. Fusion} {\bf 51} 124042

\bibitem{dlsd} Dieckmann M E, Lerche I, Shukla P K and Drury L O C 2007 {\it New J. Phys.} {\bf 9} 10

\bibitem{rom} Romanov D V, Bychenkov V Y, Rozmus W, Capjack C E and Fedosejevs R 2004 {\it Phys. Rev. Lett.} {\bf 93} 215004

\bibitem{mar1} Lazar M, Schlickeiser R and Shukla P K 2006 {\it Phys. Plasmas} {\bf 13} 102107

\bibitem{eas2} Eastwood J W 1991 {\it Comput. Phys. Comm.} {\bf 64} 252

\bibitem{dav02} Davidson R C, Hammer D A, Haber I and Wagner C E 1972 {\it Phys. Fluids} {\bf 15} 317

\bibitem{cal02} Califano F, Cecchi T and Chiuderi C 2002 {\it Phys. Plasmas} {\bf 9} 451

\bibitem{row07} Rowlands G, Dieckmann M E and Shukla P K 2007 {\it New J. Phys.} {\bf 9} 247



%\bibitem{gru} Gruzinov A 2001 {\it Astrophys. J.} {\bf 563} L15

%\bibitem{oka} Okabe N and Hattori M 2003 {\it Astrophys. J.} {\bf 599} 964



%\bibitem{sch3} Schlickeiser R 2005 {\it Plasma Phys. Control. Fusion} {\bf 47} A205

%\bibitem{fuj} Fujita Y and Kato T N 2005 {\it Mon. Not. R. Astron. Soc.} {\bf 364} 247 

%\bibitem{med1} Medvedev M V, Silva L O and Kamionkowski M 2006 {\it Astrophys. J.} {\bf 642} L1 



%\bibitem{wax} Waxman E 2006 {\it Plasma Phys. Control. Fusion} {\bf 48} B137

%\bibitem{spit} Spitkovsky A 2008 {\it Astrophys. J.} {\bf 673} L39

%\bibitem{med3} Medvedev M V 2007 {\it Astrophys. and Space Science} {\bf 307} 245

%\bibitem{yan} Yang Y T B, Gallant Y, Arons J and Langdon A B 1993 {\it Phys. Fluids B} {\bf 5} 3369

%\bibitem{tab} Tabak M, Callahan-Miller D, Ho D D-M and Zimmerman G B 1998 {\it Nucl. Fusion} {\bf 38} 509

\end{thebibliography}
\end{document}